# Attaque algébrique de NTRU à l'aide des vecteurs de Witt

GERALD BOURGEOIS   -  05 06 06  -

.

**Résumé :** on améliore une attaque algébrique de NTRU (où le paramètre $q$ est une puissance de 2), due à Silverman, Smart et Vercauteren ; au lieu de considérer, comme les précédents, les 2 premiers bits d'un vecteur de Witt attaché à la recherche de la clé secrète, on considère ici les 4 premiers bits, ce qui fournit des équations supplémentaires de degrés 4 puis 8.
   L'adjonction du $3^{\text{ème}}$ bit accélère la résolution mais celle du $4^{\text{ème}}$ bit engendre des équations de très grandes tailles.
    Cependant cette attaque ne permet pas de retrouver la clé pour des valeurs standard des paramètres.

## I- Introduction :

On présente une attaque algébrique de NTRU qui est en fait une amélioration de celle décrite dans [1] par Silverman, Smart, Vercauteren.

On considère un cryptosystème NTRU dans $\mathbb{Z}_q[X]/(X^n-1)$ où $q$ est une puissance de 2 et où $N$ est un nombre premier ; à l'aide de la méthode des vecteurs de Witt il est montré dans le papier sus mentionné que la recherche de la clé secrète se ramène à la résolution de $N$ équations quadratiques à $N$ inconnues dans $\mathbb{F}_2$ ; pour ce faire les auteurs utilisent les 2 premiers bits des vecteurs de Witt considérés.

Les équations obtenues présentent des symétries troublantes mais la méthode des bases de Gröbner n'arrive pas à mettre en évidence des propriétés algébriques qui permettraient une résolution de ce système plus rapide que celle d'un système de même taille choisi aléatoirement.

Ici on utilise les $3^{\text{ème}}$ et $4^{\text{ème}}$ bit des vecteurs de Witt ; on obtient ainsi $2N$ équations supplémentaires de degré 4 pour le $3^{\text{ème}}$ bit et de degré 8 pour le $4^{\text{ème}}$ bit ; on dispose donc d'un système de $3N$ équations à $N$ inconnues dans $\mathbb{F}_2$.

Cependant les $N$ dernières équations de degré 8 ont une telle taille que leur adjonction n'apporte rien en vitesse de calcul pour les valeurs de $N$ testées du moins avec le logiciel dont nous disposions; si on se contente des $2N$ premières équations, alors par rapport au système réduit aux $N$ premières équations, les temps de calcul sont divisés par 3 pour $N=23$ ; cependant l'écart se creuse quand $N$ augmente ; il serait intéressant d'effectuer des tests sur des valeurs de $N$ de l'ordre de 35 par exemple pour avoir une idée plus précise du gain apporté.

Remarque : les 2 attaques classiques de NTRU ( « meet in the middle » et LLL) visent à trouver une des clés de substitution à la clé secrète choisie au départ ; ici on est condamné à retrouver cette dernière clé. Pourrait on trouver une adaptation de cette méthode qui fournirait une clé de substitution ?

## II- Rappels sur les clés utilisées dans le cryptosystème NTRU :

On utilise la variante suivante de NTRU :

$N$ un nombre premier (par exemple $N=251$), $q$ une puissance de 2 (par exemple $q=128$), $p=2+X$ sont des paramètres publics.

$\mathbb{Z}_q$ désigne les entiers modulo $q$.

La clé secrète est formée par $F$ et $g \in \mathbb{F}_2[X]$, 2 polynômes de degré $N$-1 ; on peut aussi fixer le nombre de coefficients non nuls de $F$ et $g$, ce qui fournit quelques équations supplémentaires, mais nous n'en tiendrons pas compte ici.

On pose $f=1+p*F$ où $*$ est la multiplication des polynômes dans $\mathbb{Z}_q[X]/(X^n-1)$.

La clé publique est alors $h = p * f_q^{-1} * g \mod q$, où $f_q^{-1}$ est l'inverse de $f$ dans $\mathbb{Z}_q[X]/(X^n-1)$ ; cet inverse existe « toujours » si F est choisi au hasard. Les calculs d'inverse ou de multiplications $*$ se font en temps polynomial.

## III- Algébrisation de la recherche de la clé secrète $F$ :

*1°) L'isomorphisme de Witt :*

On part des nombres 2-adiques et on utilise l'isomorphisme d'anneau de Witt tronqué aux 4 premiers termes:

$$a = [a_0, a_1, a_2, a_3] \in W_4[\mathbb{F}_2] \to \sum_{i=0}^{3} a_i 2^i \mod 2^4 \in \mathbb{Z}_{2^4}, \, W_4[\mathbb{F}_2]$$ étant muni par transport de structure de ces 2 opérations :

La somme $S(a,b)$ est définie par : $S_0(a,b) = a_0 + b_0$, $S_1(a,b) = a_0 b_0 + a_1 + b_1$,

$S_2(a,b) = a_0 b_0 (a_1 + b_1) + a_1 b_1 + a_2 + b_2$,

$S_3(a,b) = a_0 b_0 a_1 a_2 + a_0 b_0 a_1 b_2 + a_0 b_0 b_1 a_2 + a_0 b_0 b_1 b_2 + a_1 b_1 a_2 + a_1 b_1 b_2 + a_2 b_2 + a_3 + b_3$.

Le produit $P(a,b)$ est défini par : $P_0(a,b) = a_0 b_0$, $P_1(a,b) = a_0 b_1 + a_1 b_0$,

$P_2(a,b) = a_0 b_0 a_1 b_1 + a_0 b_2 + a_1 b_1 + a_2 b_0$,

$P_3(a,b) = a_0 b_0 a_1 b_1 a_2 + a_0 b_0 a_1 b_1 b_2 + a_0 b_0 a_1 b_1 + a_0 b_0 a_2 b_2 + a_0 a_1 b_1 b_2 + b_0 a_1 b_1 a_2 + a_0 b_3 + b_0 a_3 + a_1 b_2 + b_1 a_2$

Notons que les choses se passent beaucoup moins bien dans $\mathbb{F}_s$ pour $s \geq 3$.

*2°) Mise en place des équations algébriques :*

On réécrit l'égalité $f * h = p * g \mod 2^4$ dans $W_4[\mathbb{F}_2]$ en posant :

$g = \sum_{i=0}^{N-1} g_i X^i$ où $g_i \in \mathbb{F}_2$, $F = \sum_{i=0}^{N-1} F_i X^i$ où $F_i \in \mathbb{F}_2$, $h = \sum_{i=0}^{N-1} h_i X^i$ où $h_i$ est connu et s'écrit dans $W_4[\mathbb{F}_2]$ : $[h_{i0}, h_{i1}, h_{i2}, h_{i3}]$.

Les inconnues sont les $(g_i)$ et les $(F_i)$ mais on ne s'intéresse qu'aux $(F_i)$.

Ainsi $f = \sum_{i=0}^{N-1} f_i X^i$ où $f_i$ s'écrit dans $W_4[\mathbb{F}_2]$:

$[1+F_{N-1}, F_0 + F_{N-1}, F_0 F_{N-1}, 0]$ si $i = 0$ et $[F_{i-1}, F_i, 0, 0]$ si $i \geq 1$.

De même $p * g = \sum_{i=0}^{N-1} R_i X^i$ où $R_i$ s'écrit dans $W_4[\mathbb{F}_2]$: $[g_{i-1}, g_i, 0, 0]$.

Enfin $f * h = \sum_{k=0}^{N-1} L_k X^k$ avec $L_k = \sum_{i+j=k \mod N} f_i h_j$.

Dans [1] on utilise les égalités $L_{k0} = R_{k0}$ et $L_{k1} = R_{k1}$ dans $W_4[\mathbb{F}_2]$ ; ici on va ajouter les égalités : $L_{k2} = R_{k2}$ et $L_{k3} = R_{k3}$ soit $L_{k2} = 0$ et $L_{k3} = 0$.

3°) *Expressions de $S_0, S_1, S_2, S_3$ dans le cas d'une somme de plus de 2 termes :*

$L_k$ est la somme de *N* termes, chaque terme étant le produit de 2 termes.
On est donc amené à calculer $S(a_1,...,a_s)$ dans $\mathbb{F}_2$ :

$$S_0(a_1,...,a_s) = \sum_i a_{i0}, \ S_1(a_1,...,a_s) = \sum_{i<j} a_{i0}a_{j0} + \sum_i a_{i1},$$

Dans $S_2$ on trouve un terme de degré 4 pour $s \geq 4$ :

$$S_2(a_1,...,a_s) = \sum_i a_{i1} \sum_{i<j} a_{i0}a_{j0} + \sum_{i<j} a_{i1}a_{j1} + \sum_i a_{i2} + \sum_{i<j<k<l} a_{i0}a_{j0}a_{k0}a_{l0}.$$

Le calcul de $S_3$ est délicat car on y trouve des termes de degré 5,6,7,8 pour $s \geq 3$, $s \geq 4$, $s \geq 6$ et $s \geq 8$ :

$$S_3(a_1,...,a_s) = \sum_i a_{i3} + \sum_{i<j} a_{i2}a_{j2} + (\sum_i a_{i2})(\sum_{i<j} a_{i1}a_{j1}) + (\sum_{i<j} a_{i0}a_{j0})(\sum_i a_{i1})(\sum_i a_{i2})$$

$$+(\sum_{i<j} a_{i0}a_{j0})(\sum_{i<j<k} a_{i1}a_{j1}a_{k1}) + (\sum_{i<j<k<l} a_{i0}a_{j0}a_{k0}a_{l0})(\sum_i a_{i2} + \sum_{i<j} a_{i1}a_{j1}) + \sum_{i<j<k<l} a_{i1}a_{j1}a_{k1}a_{l1}$$

$$(\sum_i a_{i1})(\sum_{i_1<i_2<i_3<i_4<i_5<i_6} a_{i_10}a_{i_20}a_{i_30}a_{i_40}a_{i_50}a_{i_60}) + \sum_{i_1<i_2<i_3<i_4<i_5<i_6<i_7<i_8} a_{i_10}a_{i_20}a_{i_30}a_{i_40}a_{i_50}a_{i_60}a_{i_70}a_{i_80}.$$

4°) *Les N équations associées au $3^{ème}$ bit :*

Les indices du type $i,i^*$ sont liés par $i + i^* = k \mod N$.
.
Les relations $L_{k2} = 0$ s'écrivent dans $W_4[\mathbb{F}_2]$ :

$$\sum_i (f_{i0}h_{i^*1} + f_{i1}h_{i^*0}) \sum_{i<s} f_{i0}h_{i^*0}f_{s0}h_{s^*0} + \sum_{i<s}(f_{i0}h_{i^*1} + f_{i1}h_{i^*0})(f_{s0}h_{s^*1} + f_{s1}h_{s^*0})$$

$$+\sum_i (f_{i0}h_{i^*0}f_{i1}h_{i^*1} + f_{i0}h_{i^*2} + f_{i1}h_{i^*1} + f_{i2}h_{i^*0}) + \sum_{i<j<s<t} f_{i0}h_{i^*0}f_{j0}h_{j^*0}f_{s0}h_{s^*0}f_{t0}h_{t^*0} = 0.$$

Dans [1] les relations $L_k = R_k$ pour les 2 premiers bits ont été écrites explicitement en fonction des $F_i$, ce qui permet de mettre en évidence une certaine symétrie dans les équations qui sont, rappelons le, quadratiques. Cependant cette propriété ne semble pas apporter grand-chose en terme de vitesse de calcul.

Ici nous prenons le parti de faire effectuer ces calculs par l'ordinateur :
en les $F_i$, nous obtenons un premier bloc de degré 3, des $2^{ème}$ et $3^{ème}$ blocs de degré 2 et un dernier bloc (qui est une somme de $\binom{N}{4}$ termes) de degré 4. Ainsi ces *N* équations sont de degré 4.

5°) *Les N équations associées au $4^{ème}$ bit :*

Les relations $L_{k3} = 0$ s'écrivent dans $W_4[\mathbb{F}_2]$ :

$$\sum_i [f_{i0}h_{i^*0}(f_{i1}h_{i^*1}f_{i2} + f_{i1}h_{i^*1}h_{i^*2} + f_{i1}h_{i^*1} + f_{i2}h_{i^*2}) + f_{i1}h_{i^*1}(f_{i0}h_{i^*2} + f_{i2}h_{i^*0}) + f_{i0}h_{i^*3} + f_{i1}h_{i^*2} + f_{i2}h_{i^*1}]$$

$$+\sum_{i<j}(f_{i0}h_{i^*0}f_{i1}h_{i^*1} + f_{i0}h_{i^*2} + f_{i1}h_{i^*1} + f_{i2}h_{i^*0})(f_{j0}h_{j^*0}f_{j1}h_{j^*1} + f_{j0}h_{j^*2} + f_{j1}h_{j^*1} + f_{j2}h_{j^*0})$$

$$+C_kA_k + B_kD_k \ \ C_k + B_k \sum_{i<j<s} (f_{i0}h_{i^*1} + f_{i1}h_{i^*0})(f_{j0}h_{j^*1} + f_{j1}h_{j^*0})(f_{s0}h_{s^*1} + f_{s1}h_{s^*0})$$

$$+(C_k + A_k) \sum_{i<j<s<t} f_{i0}h_{i^*0}f_{j0}h_{j^*0}f_{s0}h_{s^*0}f_{t0}h_{t^*0}$$

$$+ \sum_{i<j<s<t} (f_{i0}h_{i*1} + f_{i1}h_{i*0})(f_{j0}h_{j*1} + f_{j1}h_{j*0})(f_{s0}h_{s*1} + f_{s1}h_{s*0})(f_{t0}h_{t*1} + f_{t1}h_{t*0})$$

$$+ D_k ( \sum_{i<j<s<t<u<v} f_{i0}h_{i*0}f_{j0}h_{j*0}f_{s0}h_{s*0}f_{t0}h_{t*0}f_{u0}h_{u*0}f_{v0}h_{v*0}) .$$

$$\sum_{i_1<i_2<i_3<i_4<i_5<i_6<i_7<i_8} f_{i_10}h_{i_1*0}f_{i_20}h_{i_2*0}f_{i_30}h_{i_3*0}f_{i_40}h_{i_4*0}f_{i_50}h_{i_5*0}f_{i_60}h_{i_6*0}f_{i_70}h_{i_7*0}f_{i_80}h_{i_8*0} = 0$$

où $\quad A_k = \sum_{i<j} (f_{i0}h_{i*1} + f_{i1}h_{i*0})(f_{j0}h_{j*1} + f_{j1}h_{j*0})$, $B_k = \sum_{i<j} f_{i0}h_{i*0}f_{j0}h_{j*0}$,

$C_k = \sum_i (f_{i0}h_{i*0}f_{i1}h_{i*1} + f_{i0}h_{i*2} + f_{i2}h_{i*0} + f_{i1}h_{i*1})$, $D_k = \sum_i (f_{i0}h_{i*1} + f_{i1}h_{i*0})$.

Cette fois nous obtenons $N$ équations de degré 8 en les $F_i$.

### IV- Les tests sur 3 bits :

La machine utilisée est un portable doté d'un pentium 4 à 1,6 GHz, de 768 MO de RAM et d'un disque dur de 30 GO.

La résolution des systèmes se fait par la méthode des bases de Gröbner ; on utilise le logiciel FGb de J.C. Faugère compatible avec Maple 9.5 (*cf.* [2]).

Ce point de vue est préférable à celui exposé dans [1] où l'utilisation de XL ou de ses variantes XLS et FXL ( *cf.* [3]) était privilégiée, cela pour les raisons suivantes :
d'une part il est prouvé dans [4] que les méthodes utilisant la théorie de Buchberger sont un peu plus efficaces que XL et d'autre part il existe des implémentations de recherche de bases de Gröbner ( F5 par exemple) qui n'ont pas d'équivalent chez XL.

On a choisi $N=23$, $q=128$ et pris $F$ et $g$ au hasard de degré 22 sans fixer le nombre de termes non nuls.

Le système quadratique obtenu dans [1] à l'aide de 2 premiers bits admet entre 1 et 5 solutions et est résolu en moyenne en 4 h 46', ce qui est à peu près le temps de résolution d'un système de même taille choisi au hasard. Bien entendu il est facile de récupérer la solution qui reconstitue la clé de NTRU.

Si on tient compte du 3$^{\text{ème}}$ bit, les 23 équations supplémentaires ( $L_{k2} = 0$) sont développées en fonction des $(F_i)$ en environ 1'20".

On dispose maintenant d'un système de 46 équations à 23 inconnues ; il n'a qu'une solution qui est obtenue en moyenne en 1 h 38'. Pour $N=23$ le temps de calcul est donc divisé par 3 environ. Il semble que ce dernier facteur augmente avec $N$ ; des essais sur le système obtenu dans [1] ont été menés jusqu'à $N=35$ à l'aide du logiciel F5 de J.C. Faugère. Il serait intéressant de comparer les temps de calcul pour ces valeurs de $N$.

### V- Les tests sur 4 bits :

On se cantonne à $N=17$ pour des raisons qui apparaissent plus bas.

On tient maintenant compte du 4$^{\text{ème}}$ bit ce qui fournit encore $N$ équations supplémentaires ( $L_{k3} = 0$) cette fois de degré 8 ; cependant le développement de ces équations en fonction des $(F_i)$ est lent et l'équation réduite comporte beaucoup de termes ; ceci entraîne ensuite que la résolution du système de 51 équations à 17 inconnues (qui ne comporte aussi qu'une solution) est plus longue que celle du système intermédiaire composé de 34 équations à 17 inconnues : la raison en est que Maple prend beaucoup de temps à traduire les équations dans le langage utilisé par FGb.

Ceci est résumé dans le tableau ci dessous où les temps (ou nombres de termes) indiqués sont les temps (ou nombres de termes) maxima constatés ; remarquons que FGb peut déclarer forfait si les équations $L_{k3} = 0$ comportent environ 5700 termes.

Tests effectués avec $N=17$, $q=128$ :

|  | $L_{k2}$ | $L_{k3}$ |
|---|---|---|
| Temps de formation des $N$ équations | 84" | 1946" |
| Nombre de termes par équation réduite | 907 | 10820 |
| Temps mis par FGb | 107" ($2N$ équations) | Dépassement dans to_do ($3N$ équations) |

   Les résultats sont donc décevants car le temps de formation des équations peut être supérieur au temps de résolution du système ! Il est cependant possible qu'il existe des symétries dans les équations supplémentaires comme il en existait dans les premières. Leur connaissance permettrait de former ces équations plus rapidement mais ceci n'accélérerait sans doute pas la résolution du système qui, rappelons le, comporte des équations de degré 8.
   La méthode de Gröbner permet de se ramener à un système en gros triangulaire mais au prix d'une montée du degré des équations : les équations de degré 2 sont d'abord considérées puis lorsque, dans la résolution, le degré arrive à 4 les équations de degré 4 sont prises en compte ; enfin lorsque le degré atteint 8 les dernières équations sont utilisées.

**VI- Conclusion :**
   En considérant le $3^{\text{ème}}$ bit des vecteurs de Witt on obtient donc $N$ équations supplémentaires malheureusement de degré 4, liant les inconnues ; le système devient ainsi surdéterminé mais de façon encore insuffisante pour être résolu si l'on choisit des valeurs de $N$ considérées comme standard pour NTRU. En effet le gain de vitesse de résolution obtenu ne permet, à temps constant, qu'au mieux de passer de $N$ à $N+2$.
   L'utilisation d'un logiciel plus performant que FGb (par exemple F5) permettrait sans doute d'avoir des renseignements plus précis sur le gain (en temps de calcul) obtenu en doublant le nombre d'équations.
   On a vu que l'utilisation du $4^{\text{ème}}$ bit conduisait à des équations de degré 8 comportant, circonstance aggravante, de nombreux termes ; les temps de calcul sont ainsi augmentés du moins si on utilise FGb à travers Maple.
   D'après une conjecture de N. Courtois, il faudrait disposer de l'ordre de $N^2$ équations à $N$ inconnues pour que la résolution d'un tel système se fasse en temps polynomial. Ici cela reviendrait à écrire toutes les égalités entre composantes des vecteurs de Witt ; c'est évidemment sans espoir pour les raisons qui précèdent ; signalons aussi que la complexité des formules de Witt pour l'addition et la multiplication augmente exponentiellement avec $N$ ; par exemple l'écriture de $P_6(a,b)$ nécessite 2 feuilles Maple !

G. Bourgeois, Département de Mathématiques, Faculté de Luminy, 163 avenue de Luminy, case 901, 13288 Marseille CEDEX 09, France.
E-Mail address: bourgeoi@lumimath.univ-mars.fr